\documentclass[manuscript]{aastex}
\usepackage{amsmath}
\usepackage{lscape}

\shorttitle{shorttitle}
\shortauthors{Oya et al.}

\title{Geometric and Kinematic Structure of the Outflow/Envelope System of L1527 Revealed by Subarcsecond-resolution Observation of CS}
\author{Yoko Oya\altaffilmark{1}, Nami Sakai\altaffilmark{2}, Bertrand Lefloch\altaffilmark{3, 4}, Ana L\'{o}pez--Sepulcre\altaffilmark{1}, \\Yoshimasa Watanabe\altaffilmark{1}, Cecilia Ceccarelli\altaffilmark{3, 4}, Satoshi Yamamoto\altaffilmark{1}} 

\email{oya@taurus.phys.s.u-tokyo.ac.jp}
\altaffiltext{1}{Department of Physics, The University of Tokyo, 7-3-1, Hongo, Bunkyo-ku, Tokyo 113-0033, Japan}
\altaffiltext{2}{The Institute of Physical and Chemical Research (RIKEN), Wako, Saitama 351-0198, Japan}
\altaffiltext{3}{Universite Grenoble Alpes, IPAG, F-38000 Grenoble, France}
\altaffiltext{4}{CNRS, IPAG, F-38000 Grenoble, France}

\begin{abstract}
Subarcsecond-resolution images of the rotational line emissions of CS and c-C$_3$H$_2$ obtained toward the low-mass protostar IRAS 04368+2557 in L1527 with the Atacama Large Millimeter/submillimeter Array are investigated to constrain the orientation of the outflow/envelope system. 
The distribution of CS consists of an envelope component extending from north to south and 
a faint butterfly-shaped outflow component. 
The kinematic structure of the envelope is well reproduced by a simple ballistic model of an infalling rotating envelope. 
Although the envelope has a nearly edge-on configuration, 
the inclination angle of the rotation axis from the plane of the sky is found to be $5\degr$, where 
we find that the western side of the envelope faces the observer. 
This configuration is opposite to the direction of the large-scale ($\sim 10^4$ AU) outflow 
suggested previously from the $^{12}$CO ($J$=3--2) observation, 
and to the morphology of infrared reflection near the protostar ($\sim 200$ AU). 
The latter discrepancy could originate from high extinction by the outflow cavity of the western side, 
these discrepancies 
or may indicate that the outflow axis is not parallel to the rotation axis of the envelope. 
Position--velocity diagrams show the accelerated outflow cavity wall, 
and its kinematic structure in the $2000$ AU scale is explained by a standard parabolic model with the inclination angle derived from the analysis of the envelope. 
The different orientation of the outflow between the small and large scale 
implies a possibility of precession of the outflow axis. 
The shape and the velocity of the outflow in the vicinity of the protostar are compared with those of other protostars. 

\end{abstract}

\begin{document}
\section{Introduction}
Understanding the formation processes of rotationally supported disks around young low-mass protostars is an important target for star formation studies. 
Rotationally supported disks are usually found around low-mass Class I protostars (e.g. Hogerheijde 2001; Takakuwa et al. 2012; Yen et al. 2014), 
and a few observational evidences of disks associated with the Class 0 stage have also been reported (e.g. Yen et al. 2013; Tobin et al. 2012; Murillo et al. 2013; Ohashi et al. 2014). 
Hence, it seems likely that disk structure is formed in an early stage of protostellar evolution. 
Disk formation processes are deeply related to angular momentum 
of the infalling gas of the envelope. 
In these early phases of protostellar evolution, 
energetic outflows blow from protostars, 
which means that accretion and ejection of mass are occurring at the same time (e.g. Bachiller 1996). 
Outflows are thought to play an important role in extracting angular momentum of the infalling gas (e.g. Shu et al. 1994a, 1994b; Tomisaka 2002; Hartman 2009; Machida and Hosokawa 2013). 
Hence, both disk formation and outflow are related to angular momentum, 
and understanding one of the two would help understanding the other. 

IRAS 04368+2557 in L1527 ($d$=137 pc) is a representative Class 0 low-mass protostar, 
whose bolometric luminosity is 2.75 $L_\odot$ (Tobin et al. 2013). 
It is also known as a prototypical warm-carbon-chain-chemistry (WCCC) source (Sakai et al. 2008, 2010; Sakai and Yamamoto 2013). 
This source has a flattened infalling envelope with an edge-on configuration extending from north to south. 
Ohashi et al. (1997) and Yen et al. (2013) reported the infalling motion of the envelope gas conserving angular momentum based on interferometer observations. 
On the other hand, the existence of a Keplerian disk was suggested by Tobin et al. (2012) and Ohashi et al. (2014) by observations of the $^{13}$CO and $^{18}$CO ($J$=2--1) lines, respectively. 
Recently, Sakai et al. (2014a, 2014b) presented a clear infalling rotating motion in its envelope at a resolution of $0\farcs6$ with ALMA. 
With the aid of a simple ballistic model, 
they identified the centrifugal barrier of the infalling gas at a radius of 100 AU: 
the observed kinematic structure was well reproduced by this simple model. 
Moreover, they discovered a drastic change in chemical compositions across the centrifugal barrier (Sakai et al. 2014a). 
Carbon-chain molecules and CS mainly reside in the infalling envelope outward of the centrifugal barrier, 
whereas SO and probably CH$_3$OH are enhanced at the centrifugal barrier 
and may survive inward of it at least partly. 
Such a chemical change at the centrifugal barrier had not been anticipated before. 

In this source, molecular outflows from the protostar are extended 
toward the east-west direction which is almost perpendicular to the flattened envelope 
(i.e. almost on the plane of the sky). 
Hogerheijde et al. (1998) delineated the outflow extending over $2\arcmin$ ($\sim 2 \times 10^4$ AU) scale by observations of the $^{12}$CO ($J$=3--2) line with the James Clerk Maxwell Telescope (JCMT), 
as shown in Figure \ref{fig:previous}(a). 
The blueshifted and redshifted components are strong in the eastern side and western side of the protostar, respectively. 
For the smaller scale, Tobin et al. (2008, 2010) conducted $L^\prime$ band imaging of this source with the Gemini North telescope 
and the 3.6 $\mu$m band with the Infrared Array Camera (IRAC) on the 
$Spitzer$ Space Telescope (Figure \ref{fig:previous}(b)). 
They reported that the outflow cavity has a butterfly shape in the $10^3$--$10^4$ AU scale. 
The outflow cavity on the eastern side of the protostar is brighter than that on the western side, 
and hence they suggest that the eastern cavity would point to us. 
If so, it corresponds to the blueshifted cavity. 
This result is consistent with the orientation of the larger scale outflow observed with the $^{12}$CO line emission mentioned above. 
The protostellar-core model has so far been discussed by assuming this orientation of the outflow. 

Based on the ALMA data analysis of the envelope, 
we have fortuitously found that 
the direction of the outflow/envelope system in this source is opposite to that just described above. 
In this paper, we report the determination of the inclination angle and morphological properties of the outflow. 


\section{Results}
We used the ALMA Cycle 0 data ($\#$2011.0.00604S) reported by Sakai et al. (2014b). 
The observed lines are shown in Table \ref{tb:line}. 
The primary beam (HPBW) is $24\farcs5$. 
The synthesized beam size is 
$0\farcs8 \times 0\farcs7$ (P.A. = $-6\degr$) for the CS ($J$=5--4) and c-C$_3$H$_2$ ($5_{2, 3}$--$4_{3, 2}$) lines. 

\subsection{Overall Distribution} \label{ssec:distribution}
\begin{figure}[h]
	\includegraphics[bb = 0 0 100 400, scale = 1.0]{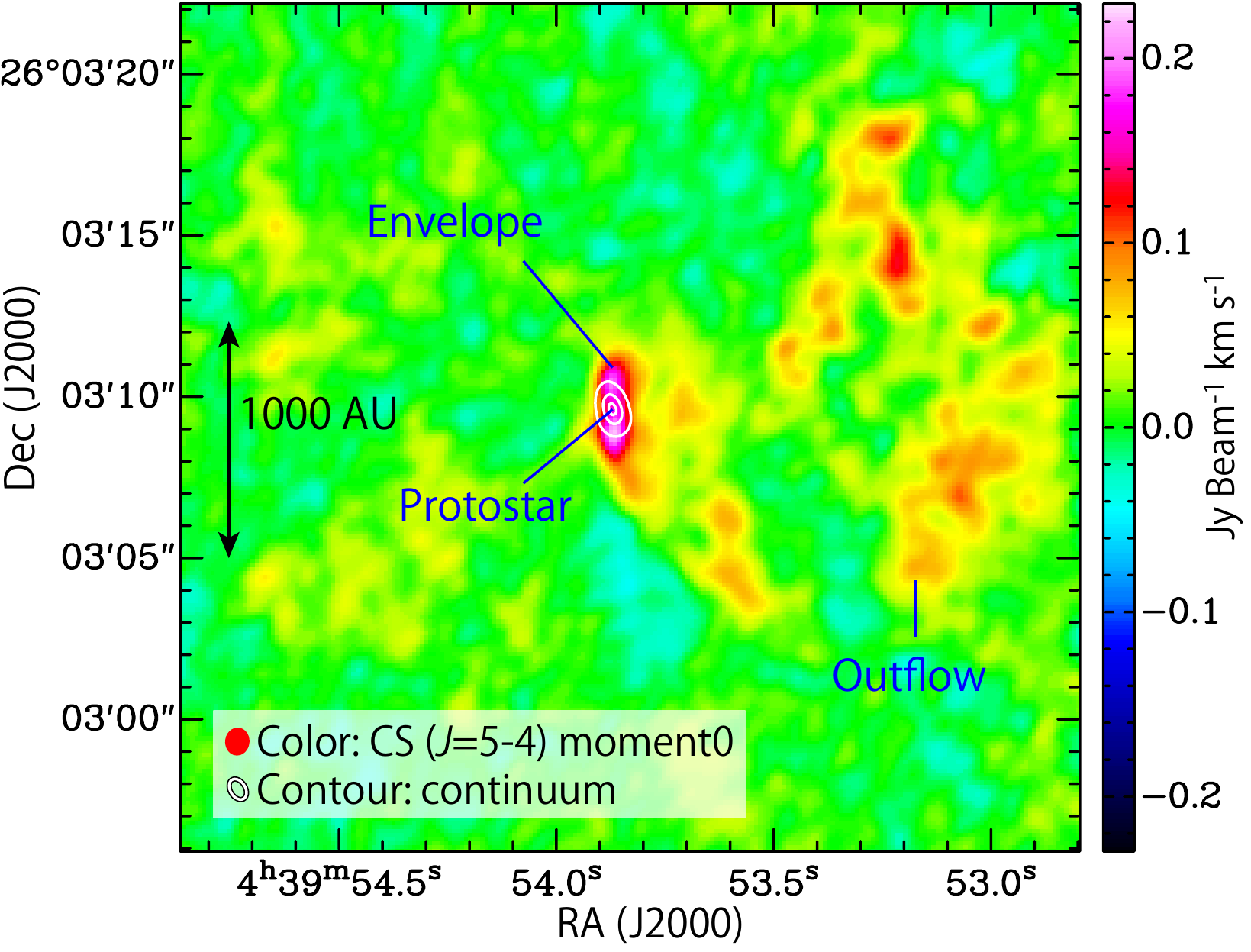}
	\caption{Moment 0 map of CS ($J$=5--4; $v_{\rm LOS} = (3.1-8.7)$ km s$^{-1}$; color) 
			and the 1.3 mm continuum map (white contours). 
			Contours for continuum are 30, 150 and 270 $\sigma$ 
				where $\sigma$ is 1 mJy beam$^{-1}$ km s$^{-1}$. 
			\label{fig:moment0}}
\end{figure}

Figure \ref{fig:moment0} shows the moment 0 map of the CS ($J$=5--4) line. 
The envelope component 
extends almost along the north-south axis, 
as reported in other lines (Sakai et al. 2014a, 2014b). 
Although the emission of the outer part is heavily resolved out, 
a part of the butterfly-shaped outflow cavity wall (Tobin et al. 2010) can be seen mainly on the western side of the protostar in addition to the envelope component. 
The high-density gas traced by the CS ($J$=5--4) line is distributed asymmetrically around the protostar. 

Figure \ref{fig:PV_PA270} shows the position-velocity (PV) diagram along the outflow axis. 
Envelope components are concentrated at the protostar position, while outflow components are extended along the east-west axis from the protostar. 
The outflow seems to be accelerated with increasing distance from the protostar. 
Although blueshifted components are prominent on the western side of the protostar, 
the velocity shifts of the outflow are almost symmetrical to the protostar position. 
It is therefore obvious that the outflow axis is close to the plane of the sky, 
as reported previously (Tobin et al. 2008, 2010). 
Unfortunately, it is difficult to derive the inclination angle accurately from the kinematic structure of the outflow, 
because the outflow components are faint and heavily resolved out. 
Hence, we first investigate the envelope components to discuss the geometry of the outflow/envelope system. 

\subsection{Envelope} \label{ssec:env}
The kinematic structure of the protostellar envelope of L1527 observed in the c-C$_3$H$_2$, CCH, and CS emission is well reproduced with a model of an infalling rotating envelope (Sakai et al. 2014a, 2014b). 
Details of this model are described in the paper investigating another low-mass protostellar source: IRAS 15398--3359 (Oya et al. 2014). 
In this model, the particles cannot fall inward of a certain radius because of conservation of angular momentum and energy, 
and this radius defines the centrifugal barrier. 
In this ballistic model, 
the velocity field of the particle motion is characterized by the protostellar mass and the radius of the centrifugal barrier. 
The PV diagram of the model prepared along the envelope reflects only the absolute value of its inclination angle, 
and it does not reflect the direction of its inclination from the plane of the sky. 
Hence, Sakai et al. (2014a) did not consider the direction of the inclination in the analysis of the PV diagram. 
Therefore, we here examine the PV diagrams of the CS ($J$=5--4) and c-C$_3$H$_2$ ($5_{2, 3}$--$4_{3, 2}$) lines 
prepared along various lines passing through the protostar position to investigate the direction of the inclination of the envelope. 

Figures \ref{fig:PV_I5} and \ref{fig:PV_I-5} show the PV diagrams of the CS ($J$=5--4) line 
along the six lines shown in Figure \ref{fig:PV_direction}(a) 
for the inclination angle of $+5\degr$ and $-5\degr$, respectively, 
where the positive and negative inclination angles correspond to the cases shown in Figures \ref{fig:PV_direction}(b) and (c), respectively. 
It should be noted that Tobin et al. (2013) assumed an inclination angle of $-5\degr$ (Figure \ref{fig:PV_direction}(c)). 
The other physical parameters assumed in the model are described in Appendix A.
The emission of CS ($J$=5--4) at the centrifugal barrier is more enhanced than in the outer part of the envelope compared with the c-C$_3$H$_2$ ($5_{2, 3}$--$4_{3, 2}$) case (Sakai et al. 2014b). 
Moreover, the redshifted components are slightly weaker than the blueshifted components, 
and the systemic velocity components are self-absorbed. 
Nevertheless, 
the kinematic structure along the envelope (``$180\degr$") is well explained by the above model 
assuming the either direction of the inclination ($+5\degr$ or $-5\degr$; Sakai et al. 2014a). 
Note that resolved-out components are negligible, 
because we focus on the compact distribution around the protostar for the envelope analyasis. 

However, it is evident that the positive inclination well reproduces the PV diagrams along all the lines (Figure \ref{fig:PV_I5}), 
whereas the negative inclination does not (Figure \ref{fig:PV_I-5}). 
In particular, 
the PV diagrams labeled as ``$240\degr$", ``$270\degr$", ``$300\degr$" in Figures \ref{fig:PV_I5} and \ref{fig:PV_I-5}, 
which reflect the infalling motion rather than the rotating motion, 
seem to be reproduced better by the model with the inclination angle of $+5\degr$ (Figure \ref{fig:PV_I5}) 
than that with the inclination angle of $-5\degr$ (Figure \ref{fig:PV_I-5}). 
For instance, the observed  PV diagram along the ``$270\degr$" line shows two peaks; 
one at the eastern side with the redshifted velocity 
and the other at the western side with the blueshifted velocity. 
In the model assuming the inclination angle of $+5\degr$, 
the peaks appear as the observational trend mentioned above. 
In the model assuming the inclination angle of $-5\degr$, 
the two peaks, on the other hand, appear in the opposite way; 
one at the eastern side with the blueshifted velocity and the other at the western side with the redshifted velocity. 
Hence, the inclination angle of $+5\degr$ better explains the observation. 

We calculated the root-mean-square (rms) of the difference between the observed and simulated PV diagrams 
in order to evaluate the goodness of the fit more quantitatively. 
However, this attempt was not very successful. 
Our model is a simplified one involving many assumptions described in Appendix A 
in order to explore just the basic physical and kinematic structure of the envelope. 
Hence, we have to note that the rms of the difference suffers from systematic errors due to these assumptions. 
Nevertheless, the rms of the difference for the PV diagram along the ``$270\degr$" line is 0.022 Jy beam$^{-1}$ for the inclination angle of $-5\degr$, 
which is confirmed to be larger than that for the inclination angle of $+5\degr$ (0.018 Jy beam$^{-1}$). 

We also attempted to optimize the inclination angle from the fit of the PV diagrams. 
For this purpose, we conducted simulations for the other inclination angles, as shown in Appendix A. 
Since the fit is not perfect due to the simplicity of the model, 
the statistical argument based on the rms of the difference is almost meaningless, as mentioned above. 
Nevertheless, we found that the inclination angle from $+5\degr$ to $+10\degr$ reasonably reproduces the observation. 
More safely, it is constrained to be less than $+15\degr$. 
In addition, we can firmly conclude that the inclination angle is positive: 
the positions and the velocities of the two peaks mentioned above are 
not consistent with the observations, 
if negative inclination angles are employed for the model. 
Tobin et al. (2008) reported that 
the morphology of the $L^\prime$ band image cannot be explained, 
if the disk is inclined by much larger than $10\degr$ (much smaller than $80\degr$ of their definition). 
Hence, our estimation of the inclination angle is consistent with theirs except for the direction of the inclination. 
In the following discussion, we employ the inclination angle of $+5\degr$. 

%

The positive inclination angle means that the envelope in front of the protostar appears in the eastern side of the protostar, 
and the outflow axis in the western side of the protostar points to us (Figure \ref{fig:PV_direction}(b)). 
This direction is opposite to that reported previously (Tobin et al. 2010; Hogerheijde et al. 1998). 
However, it is consistent with the moment 1 map of SO ($J_N$=$7_8$--$6_7$) reported by Sakai et al. (2014b). 
It shows a slightly skewed feature, 
where the redshifted component slightly extends from the northern part to the eastern part due to the rotating and infalling motion, 
and the blueshifted component slightly extends from the southern part to the western part. 
This indicates that the eastern part and the western part of the envelope are in front of and behind the protostar, respectively (Figure \ref{fig:PV_direction}(b)). 
This configuration is consistent with our interpretation. 

Figure \ref{fig:PV_c-C3H2} shows the PV diagrams of c-C$_3$H$_2$ ($5_{2, 3}$--$4_{3, 2}$) along the six lines passing through the protostar position (Figure \ref{fig:PV_direction}(a)). 
Although the redshifted components are weaker than the blueshifted components and the systemic velocity components are self-absorbed as in the case of CS, 
these diagrams seem to be well reproduced by the model with the inclination angle of $+5\degr$, 
as represented by the blue contours. 
Because c-C$_3$H$_2$ preferentially exists in the envelope (Sakai et al. 2014a), 
the CS emission which is not seen in the c-C$_3$H$_2$ emission would trace the outflow component. 

\subsection{Outflow} \label{ssec:outflow}
\begin{figure}[h]
	\includegraphics[bb = 0 0 100 400, scale = 1.1]{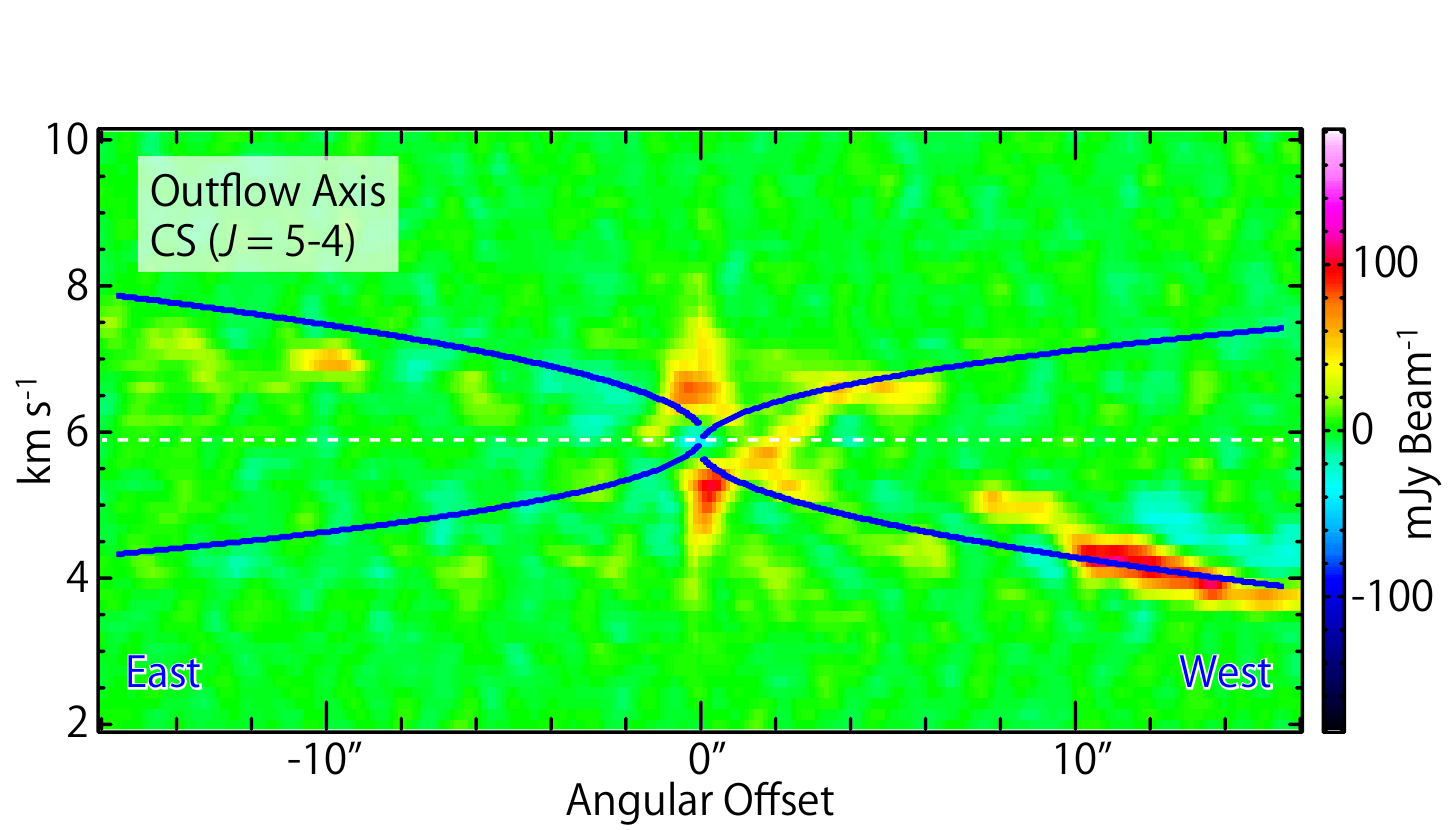}
	\caption{PV diagram of CS ($J$=5--4; color) along the outflow axis. 
			The blue lines represent the result of the outflow model with the parameters of 
			$I = +5\degr$, $C = 0.05$ arcsec$^{-1}$ and $v_0 = 0.10\ {\rm km\ s}^{-1}$, 
			where no offset of the origin of the outflow from the protostar is assumed in this figure. 
			\label{fig:PV_PA270}}
\end{figure}

On the basis of the kinematics of the envelope, 
we constrained the inclination angle of the envelope. 
Then we investigate the outflow structure seen in Figure \ref{fig:PV_PA270}, 
assuming that the outflow axis is perpendicular to the mid-plane of the infalling rotating envelope. 
We employ a standard model of the outflow structure for the analysis (Lee et al. 2000; Oya et al. 2014). 
In this model, 
the shape of the outflow cavity wall and the velocity field on the wall are approximated by the following formulae (Lee et al. 2000): 
\begin{equation}
	z = C R^2, \quad v_R = v_0 \frac{R}{R_0}, \quad v_z = v_0 \frac{z}{z_0}, \label{eq:outflow}
\end{equation}
where the $z$ axis is taken along the outflow axis with an origin at the protostar, 
and $R$ denotes the radial size of the cavity perpendicular to $z$-axis. 
$R_0$ and $z_0$ are normalization constants, and both are set to be $1^{\prime \prime}$ (Oya et al. 2014).  
$C$ and $v_0$ are free parameters. 
Thus, the wall of the outflow cavity has a parabolic shape, 
and it is linearly accelerated with increasing distance from the protostar along the outflow axis and with increasing distance from the outflow axis. 
Such a parabolic model is widely applied to various low-mass and high-mass protostellar sources 
(e.g. Arce et al. 2013; Beuther et al. 2004; Lumbreras \& Zapata 2014; Takahashi \& Ho 2012; Takahashi et al. 2013; Yeh et al. 2008; Zapata et al. 2014). 

The blue lines shown in Figure \ref{fig:PV_PA270} represent the best-fit result by eye, 
where the inclination angle is fixed to $+5\degr$, which is derived from the kinematic structure in the envelope. 
The parameters are: 
$C = 0.05\ {\rm arcsec}^{-1}$ and $v_0 = 0.10\ {\rm km\ s}^{-1}$. 
The accelerated outflow component in the western side of the protostar seems to be explained by this parameter. 
However, the velocity structure in the vicinity of the protostar is not well reproduced. 
This result suggests that the origin of the parabolic shape has a certain offset from the protostar position, 
as pointed out by Tobin et al. (2008). 

\begin{figure}[h]
	\includegraphics[bb = 0 0 100 300, scale = 0.85]{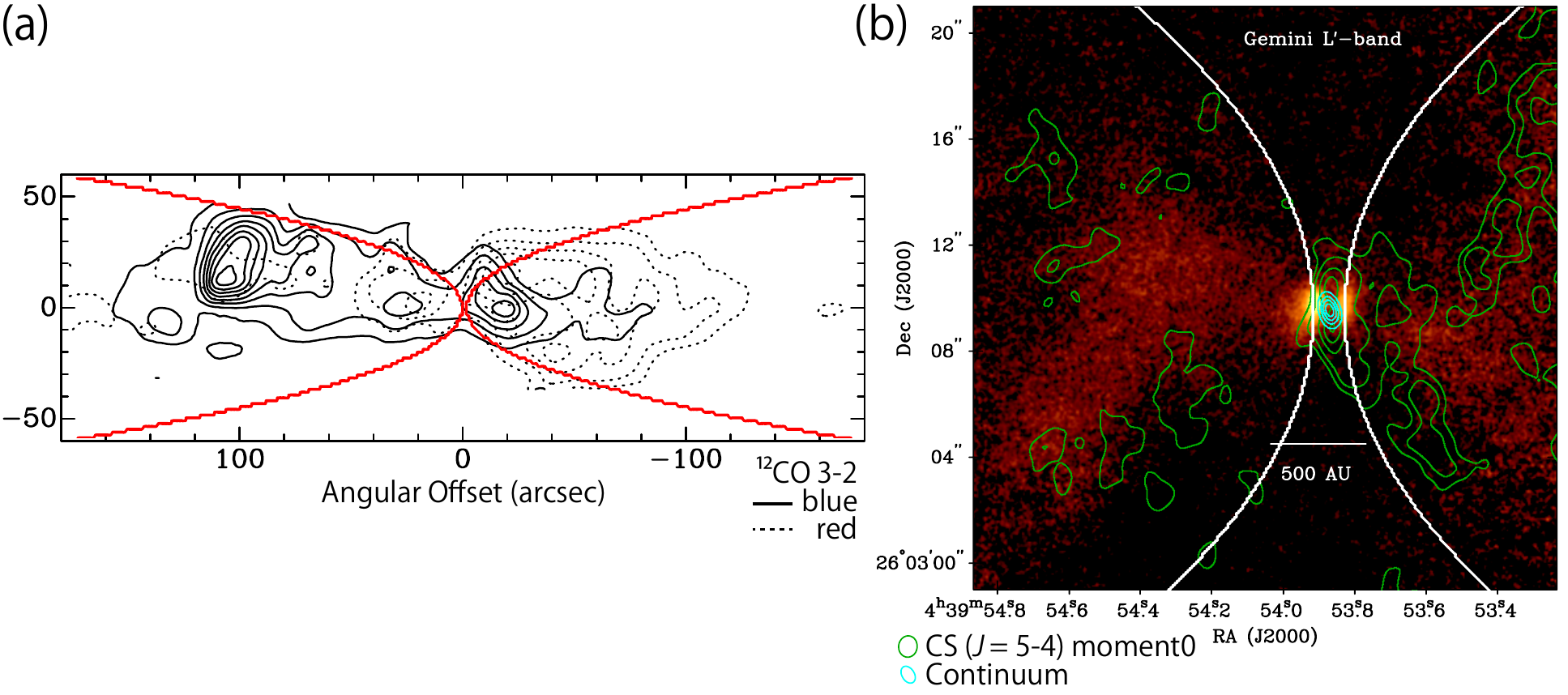}
	\caption{(a) $^{12}$CO map (black contours) by MacLeod et al. (1994) (Hogerheijde et al. 1998). 
				The FWHM beam size is $11\arcsec$ and 
				the contours are drawn at the 3 $\sigma$ level. 
			(b) $L^\prime$ band observation (color) by Tobin et al. (2010), 
				the moment 0 map of CS ($J$=5--4; green contours)
				and the 1.3 mm continuum map (light blue contours). 
				Contours for CS are 3, 6, 12, 18, and 24 $\sigma$, 
				where $\sigma$ is 9 mJy beam$^{-1}$ km s$^{-1}$. 
				Contours for the continuum are every 20 \% of the peak intensity, 
				which is 308 mJy beam$^{-1}$ km s$^{-1}$. 
			Red lines in the panel (a) and white lines in the panel (b) represent 
				the best-fit model of the outflow, 
				where the parameters are: $I = +5\degr$, $C = 0.05\ {\rm arcsec}^{-1}$ 
				and $v_0 = 0.10\ {\rm km\ s}^{-1}$, 
				and the origins of the lobes have an offset of $0\farcs62$ from the protostar position (Tobin et al. 2010). 
			\label{fig:previous}}
\end{figure}

\begin{figure}[h]
	\includegraphics[bb = 0 0 100 1100, scale = 0.37]{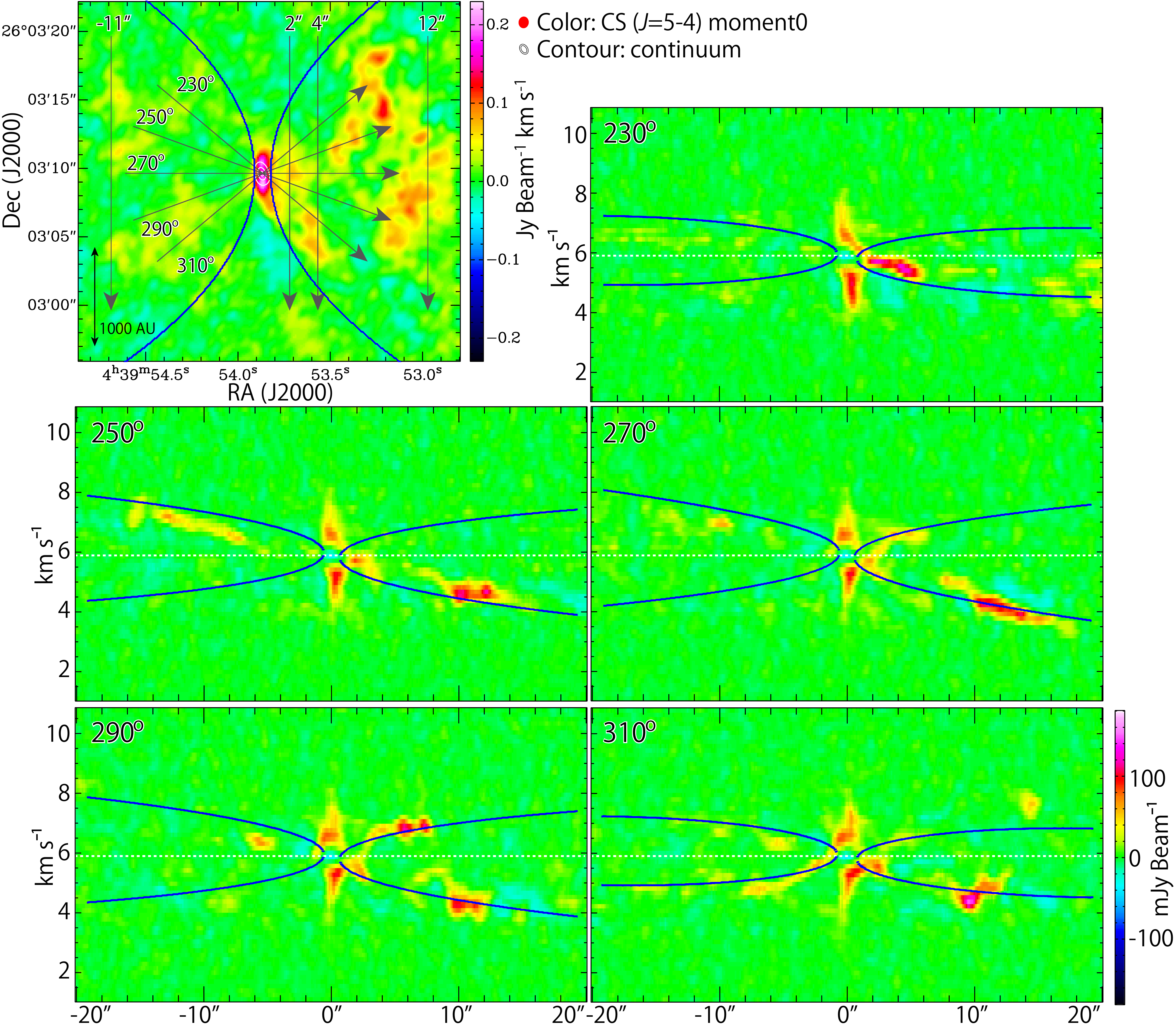}
	\caption{Moment 0 map (upper left panel) and PV diagrams of CS ($J$=5--4). 
			The moment 0 map of CS is the same as in Figure \ref{fig:moment0}. 
			White contours in the moment 0 map represent the continuum map 
			(See Figure \ref{fig:moment0} for the contour levels). 
			PV diagrams are prepared along the gray arrows 
			passing through the protostar position shown in the moment 0 map. 
			The blue lines represent the result of the best-fit outflow model, 
			where the parameters are: 
			$I = +5\degr$, $C = 0.05\ {\rm arcsec}^{-1}$ and $v_0 = 0.10\ {\rm km\ s}^{-1}$, 
			and the origins of the lobes have an offset of $0\farcs62$ from the protostar position (Tobin et al. 2010). 
			\label{fig:PV_outflow}}
\end{figure}

We therefore adopt the offset ($0\farcs62 \sim 85$ AU) of the outflow origin reported by Tobin et al. (2010) in the model. 
The upper left panel in Figure \ref{fig:PV_outflow} shows the moment 0 map of CS ($J$=5--4), 
whereas the other panels in Figure \ref{fig:PV_outflow} are the PV diagrams along the gray arrows shown in the upper left panel. 
The blue lines in the PV diagrams represent the best model with 
the inclination angle of $+5\degr$, where $C = 0.05\ {\rm arcsec}^{-1}$ and $v_0 = 0.10\ {\rm km\ s}^{-1}$. 
When the inclination angle is higher than $+15\degr$ or lower than 
$-5\degr$, 
the model does not reproduce the observations well with any values of $C$ and $v_0$. 
This constraint is consistent with that derived from the kinematic structure of the envelope (Section \ref{ssec:env}). 
The red lines in Figure \ref{fig:previous}(a), the white lines in Figure \ref{fig:previous}(b), 
and the blue lines in the upper left panel in Figure \ref{fig:PV_outflow} represent the best model with the inclination angle of $+5\degr$. 
Although the parameters for the model are derived from the kinematic structure of the outflow/envelope system, 
the model seems to explain the spatial extent of the outflow in the vicinity of the protostar. 
The maximum velocity shifts from the systemic velocity (5.9 km s$^{-1}$; Sakai et al. 2010) 
along the line of sight for the eastern and western lobe in Figure \ref{fig:previous}(a) are 
reported to be 6.9 and 9.6 km s$^{-1}$, respectively, from the $^{12}$CO ($J$=3--2) observation (Hogerheijde et al. 1998). 
In the above model, the maximum velocity shift for the two lobes is $8.2$ km s$^{-1}$ at the distance of $200^{\prime \prime}$ from the protostar. 
Then, the physical parameters for the model appear to be roughly consistent with the larger-scale observation, 
even though they are derived from the kinematic structure in 
the spatial scale of a few tens of arcsecond. 

Figure \ref{fig:PV_across} shows the PV diagrams of the outflow along the lines perpendicular to the outflow axis shown in the upper left panel in Figure \ref{fig:PV_outflow}. 
The blue lines represent the best model, 
which seems to reproduce the basic features. 
When the outflow axis is almost parallel to the plane of the sky, 
PV diagrams across the outflow axis show an elliptic feature (Oya et al. 2014). 
A part of such an elliptic feature is indeed seen in the PV diagrams. 

It should be noted that the kinematic and geometric structure might be different between the eastern and western lobes, 
because the simple model cannot perfectly explain the distribution (for instance, the PV diagrams in Figure \ref{fig:PV_outflow}). 
This may originate from different distributions of ambient gas between the two sides. 
However, 
a detailed discussion on this issue is outside the scope of this paper, since 
the CS ($J$=5--4) distributions in the outflow are faint and heavily resolved out. 

\section{Discussion}
\subsection{Direction of the outflow} \label{ssec:direction}
We evaluated the inclination angle of the infalling rotating envelope with its direction by use of a simple ballistic model. 
Assuming that the outflow axis is perpendicular to the mid-plane of the envelope, 
the direction of the inclination contradicts previous reports (Tobin et al. 2008, 2010; Hogerheijde et al. 1998). 
Here, we discuss this contradiction. 

Tobin et al. (2008, 2010) assumed a configuration in which the eastern part of the envelope faces the observer (Figure \ref{fig:PV_direction}(c)), 
because the infrared ($L^\prime$ band) reflection by the cavity wall of the outflow 
in the vicinity of the protostar is brighter in the eastern part. 
The contradiction can originate from the following two reasons. 

The first possibility is the extinction by inhomogeneous gas distribution around the protostar. 
Green contours in Figure \ref{fig:previous}(b) represent the moment 0 map of the CS ($J$=5--4) emission. 
In the southwestern side of the protostar, 
a ridge structure is extended along the western outflow-cavity, where the CS ($J$=5--4) emission is bright. 
In this ridge, the emission in the $L^\prime$ band is relatively weak. 
This indicates that the dense gas on the near side of the outflow cavity obscures the scattered light in the $L^\prime$ band. 
In fact, the upper right panel in Figure \ref{fig:PV_across} shows that the bright CS emission of this ridge is slightly blueshifted, 
which indicates that this component is in front of the outflow axis. 
We evaluated the CS column density to be (1--2) $\times 10^{13}$ cm$^{-2}$ toward the western position offset from the protostar by $0\farcs7$, 
which is the FWHM of the synthesized beam along the east-west axis. 
This value was derived using the RADEX code (van der Tak et al. 2007), 
where the H$_2$ density and the kinetic temperature were assumed to be 
$10^6$--$10^7$ cm$^{-3}$ and 30 K, respectively (Sakai et al. 2014a, 2014b). 
The H$_2$ column density is then roughly estimated to be (1--2) $\times 10^{22}$ cm$^{-2}$ 
by using a CS fractional abundance relative to H$_2$ of $10^{-9}$, 
a typical value for protostellar cores (e.g. van Dishoeck et al. 1995; Watanabe et al. 2012). 
This H$_2$ column density corresponds to an $L^\prime$ band extinction of about (0.5--1) mag. 

Considering that the observed CS emission is heavily resolved out and mainly traces dense gas ($> 10^6$ cm$^{-3}$), 
the $L^\prime$ band extinction estimated above could be the lower limit. 
If the emission in the $L^\prime$ band is weakened  
by the outflow cavity wall, 
the brightness on the two sides of the protostar depends on the distribution of matter in the outflow cavity, 
and the brightness asymmetry cannot constrain the direction of the inclination. 
Very recently, Velusamy et al. (2014) reported that the $Spitzer$ 3.6 $\mu$m HiRes deconvolved image shows 
the brightness asymmetry of the outflow cavity in the vicinity of the protostar which is opposite to that reported by Tobin et al. (2008, 2010). 
The result by Velusamy et al. (2014) seems consistent with the positive inclination angle that we found in this study. 
Thus, the observations of the morphology of the cavity reflection seem inconsistent with each other. 
Hence, the configuration of the disk/outflow system derived from the kinematics of the infalling rotating envelope is more reliable than that from the morphology. 

The second possibility solving the contradiction is that 
the outflow axis is not perpendicular to the mid-plane of the envelope. 
This may be possible due to some asymmetry of the protostellar system such as binarity, 
although the binarity in this source is controversial (Loinard et al. 2002). 
In recent studies of exoplanets, 
it is reported that the spin axis of the central star is not always parallel to the orbital axes of planets 
(e.g. Xue et al. 2014). 
Although the scattering among planets are considered as an important cause, 
the angular momentum of the central star and the disk could have different axes. 
This is an important issue for exo-planet studies. 
In this relation, exploring this possibility would be very interesting. 

According to Hogerheijde et al. (1998), the outflow seen in the $^{12}$CO ($J$=3--2) line is mainly blueshifted and redshifted in the eastern and western side of the protostar, respectively, 
in a large scale of $\sim100\arcsec$, 
although the counter velocity components are seen in both lobes. 
If the direction of the small-scale outflow is different from that of the large-scale one, 
it may imply that the outflow is precessing. 
The dynamical time scales for the eastern and western lobe are reported to be $1.5 \times 10^4$ and $9.0 \times 10^3$ yr, respectively, 
which are not corrected for inclination (Hogerheijde et al. 1998). 
Note that the dynamical time scales are rough estimates based on the apparent maximum velocity shifts and the lengths of the lobes. 
Here, it should be stressed that the velocity along the line of sight ($v_{\rm LOS}$) almost consists of the velocity component perpendicular to the outflow axis, 
which represents an expanding motion of the cavity wall, 
because of the almost edge-on geometry. 
Hence, the velocity along the outflow axis can hardly be estimated by $v_{\rm LOS} / \sin I$, 
where $I$ is the inclination angle of the outflow axis 
($0\degr$ for the case parallel to the plane of the sky). 
When the apparent distance from the protostar on the plane of the sky is $100\arcsec$, 
the distance from the protostar along the outflow axis ($z$) is also $\sim 100\arcsec$ with the inclination angle of $+5\degr$. 
The line of sight and each outflow cavity wall have two intersections, 
which have different line-of-sight velocities ($v_{\rm LOS}$). 
In the redshifted lobe, 
the velocity along the outflow axis ($v_z$) and the expanding velocity ($v_r$) 
at the distance projected on the plane of the sky of $100\arcsec$ from the protostar 
are calculated to be 
$(v_z, v_r) = (9.6\ {\rm km\ s}^{-1}, 4.4\ {\rm km\ s}^{-1})$ and $(v_z, v_r) = (10.4\ {\rm km\ s}^{-1}, 4.6\ {\rm km\ s}^{-1})$ for the two intersections. 
Hence, the velocities along the line-of-sight ($v_{\rm LOS}$) are calculated to be $-3.5$ and $5.4\ {\rm km\ s}^{-1}$. 
In the blueshifted lobe, 
the velocity along the outflow axis and the expanding velocity are the same as that in the redshifted lobe, 
while the velocity shifts from the systemic velocity along the line-of-sight ($v_{\rm LOS}$) have the opposite sign. 
Then, the age of the lobe at the distance of $100\arcsec$ from the protostar is estimated to be $t = z / v_z \sim 6.5 \times 10^3$ yr. 
Assuming that the inclination angle is $+5\degr$ in the $10\arcsec$ scale and $-5\degr$ in the $100\arcsec$ scale, 
we speculate a precession of $10\degr$ in $\sim 6.5 \times 10^3$ yr. 
Here, we assume that there was no other precession event in between. 
This is comparable to the case of L1157, 
where a precession of $6\degr$ is observed with a period of $\sim 4 \times 10^3$ yr (Gueth et al. 1996). 

An alternative possibility is that emission of the outflow cavity is not homogeneous, and the blueshifted and redshifted emissions are dominant in the large scale eastern and western lobes, respectively, by accident. 
This situation could happen because the outflow is blowing almost along the plane of the sky. 

\subsection{Angular momentum} \label{ssec:angularmomentum}
In the model of the infalling rotating envelope, 
it is assumed that the gas cannot fall inward of the centrifugal barrier 
due to angular momentum conservation (Oya et al. 2014). 
However, Sakai et al. (2014a) pointed out that H$_2$CO possibly resides inward of the centrifugal barrier. 
If so, the gas has to lose its angular momentum to fall beyond the centrifugal barrier. 
The outflow is a candidate mechanism extracting the angular momentum from the envelope gas (e.g. Tomisaka 2002; Machida and Hosokawa 2013). 
Hence, it is interesting to explore the possibility of outflow rotation. 
In fact, rotation of outflow and jet is suggested 
for various protostellar sources 
(Choi et al. 2011; Codella et al. 2007; Coffey et al. 2007; Hara et al. 2013; Launhardt et al. 2009; Lee et al. 2007, 2008, 2009; Pech et al. 2012; Zapata et al. 2010). 
Rotation is mostly observed for a collimated jet, 
while rotation of a well spread outflow like the L1527 outflow is not very evident (B59$\#$11: Hara et al. 2013). 

\begin{figure}[h]
	\includegraphics[bb = 0 0 100 800, scale = 0.38]{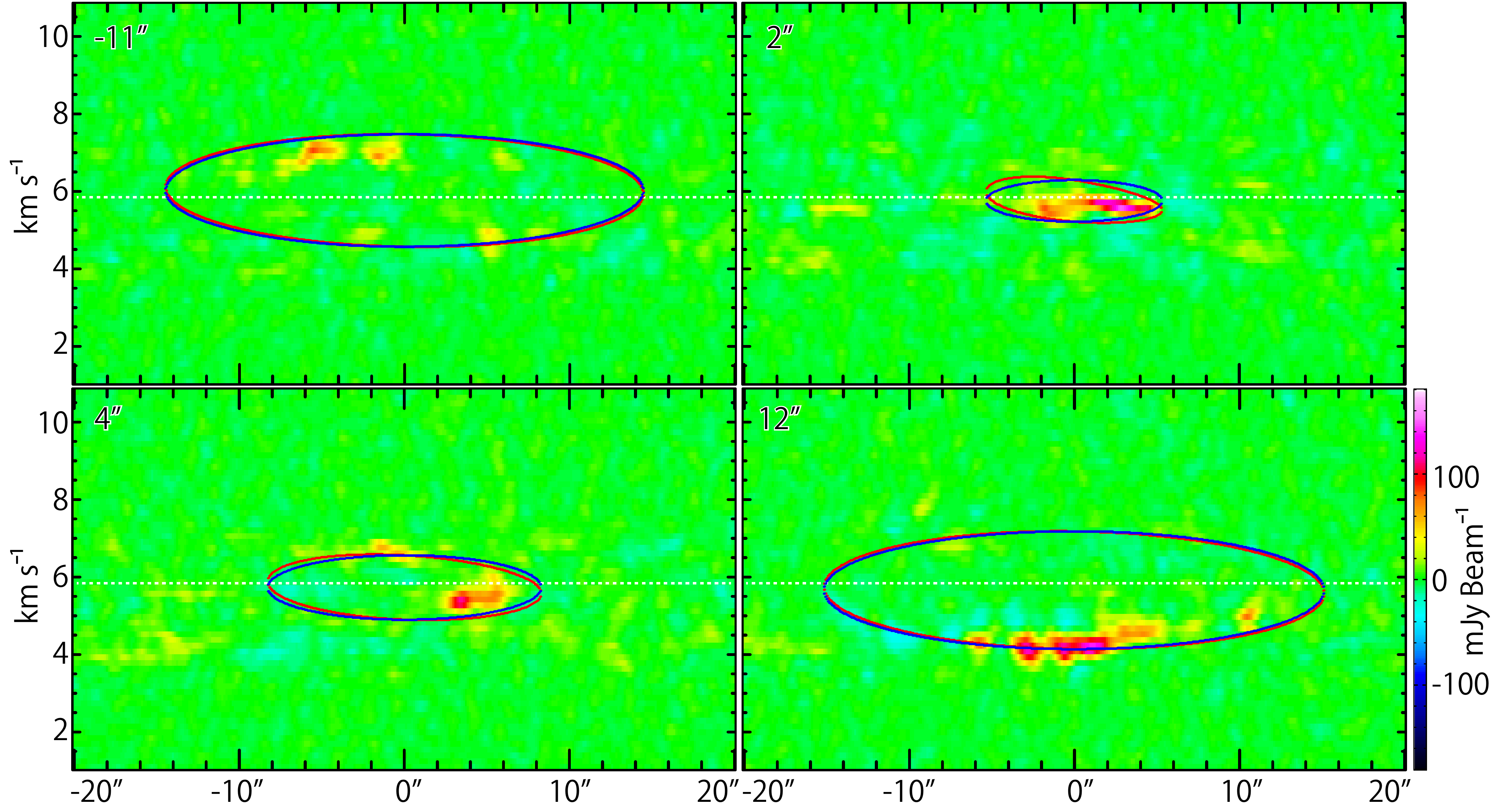}
	\caption{PV diagrams of CS ($J$=5--4; color) across the outflow axis, 
			where the position axes are shown in the upper left panel in Figure \ref{fig:PV_outflow}. 
			The blue lines represent the result of the best-fit outflow model, 
			where the parameters are: 
			$I = +5\degr$, $C = 0.05\ {\rm arcsec}^{-1}$ and $v_0 = 0.10\ {\rm km\ s}^{-1}$, 
			and the origins of the lobes have an offset of $0\farcs62$ from the protostar position (Tobin et al. 2010). 
			The red lines represent another outflow model, 
			where a rotating motion of the outflow cavity wall is taken into account (See Section \ref{ssec:angularmomentum}), 
			although they are almost overlapped with the blue lines. 
			\label{fig:PV_across}}
\end{figure}

As shown in Figure \ref{fig:PV_across}, 
the velocity field on the outflow cavity wall is almost symmetric to the outflow axis. 
A rotating motion is hardly seen in the outflow cavity wall. 
If the angular momentum in the envelope was extracted by the outflow, 
the northern edge and the southern edge of the outflow should be 
more redshifted and blueshifted than the model, respectively. 
The red lines in Figure \ref{fig:PV_across} represent another outflow model, 
where the rotating motion of the outflow cavity wall is taken into account by a simple model. 
In this rotating outflow model, it is assumed that the specific angular momentum 
on the outflow cavity wall is conserved (Launhardt et al. 2009), 
and the launching point of the outflow is at the centrifugal barrier. 
The specific angular momentum of the outflow in the model is roughly approximated to be the same as that of the envelope gas, 
which is determined by the protostellar mass and the radius of the centrifugal barrier (Section \ref{ssec:env}). 
This approximation corresponds to the maximum specific angular momentum allowed for the outflow. 
Then the rotating motion on the outflow cavity wall is represented by the radial size $R$ of the outflow cavity perpendicular to the outflow-axis as:
\begin{equation}
	v_{\rm rot} = \frac{\sqrt{2 r_{\rm CB} G M}}{R}, 
	\label{eq:rot_outflow}
\end{equation}
where $r_{\rm CB}$ denotes the radius of the centrifugal barrier in the envelope, 
$M$ denotes the protostellar mass, and $G$ is the gravitational constant (Oya et al. 2014). 
The rotating motion on the outflow cavity wall would be large near the launching point of the outflow 
(small $R$; the upper right panel in Figure \ref{fig:PV_across}), 
and decrease with increasing distance from the protostar 
(large $R$; the bottom right panel in Figure \ref{fig:PV_across}). 
This would be a reason why the rotation motion has been found in the well collimated jets. 
Figure \ref{fig:PV_across} shows that the results of the two models are almost overlapped with each other, 
even if all the angular momentum in the envelope is carried by the outflow. 
Therefore, it is essential to investigate the launching point of the outflow at a higher spatial resolution. 

\subsection{Comparison with other sources} \label{ssec:comparison}
In this paper, we derived the outflow parameters, $C$ and $v_0$, by using a standard outflow model. 
Although the CS emission from the outflow cavity wall is heavily resolved out, 
we can reasonably obtain these parameters by analyzing both the geometrical and kinematic structures simultaneously. 
Recently, we also investigated the outflow of IRAS 15398--3359, which is a low-mass protostar in the Lupus 1 molecular cloud, by using the same model (Oya et al. 2014). 
Here, we compare the outflow parameters for L1527 with those for IRAS 15398--3359. 
Because the parameters $C$ and $v_0$ in the equations (\ref{eq:outflow}) are defined as the values at $1\arcsec$, 
they should be redefined as the values at 1 AU 
to compare the outflow parameters for the sources which have different distances from the Sun. 
Then, the equations are written as 
\begin{equation}
	z_{\rm AU} = c_{\rm AU}\ r_{\rm AU}^2, \quad v_r = v_{\rm AU} \frac{r_{\rm AU}}{r_0},  \quad v_z = v_{\rm AU} \frac{z_{\rm AU}}{z_0},
	\label{eq:AU}
\end{equation}
where $z_{\rm AU}$ denotes the distance to the protostar along the outflow axis, 
and $r_{\rm AU}$ the radial size of the cavity perpendicular to $z$-axis. 
$z_0$ and $r_0$ are normalization constants, and both are set to be 1 AU. 
$c_{\rm AU}$ and $v_{\rm AU}$ are free parameters. 
The units are AU for $z_{\rm AU}$ and $r_{\rm AU}$, 
AU$^{-1}$ for $c_{\rm AU}$, and km s$^{-1}$ for $v_{\rm AU}$. 
When the source is at the distance of $D$ pc from us, 
$z_{\rm AU} = zD$, $r_{\rm AU} = RD$, $c_{\rm AU} = C / D$, 
and $v_{\rm AU} / r_0 = v_0 / R_0 D$. 
The $v_0 / R_0 D$ ($= v_{\rm AU} / r_0$) denotes the velocity perpendicular to the outflow axis at the radial size of the cavity of 1 AU, 
and $C / D$ ($= c_{\rm AU}$) denotes 
the curvature of the cavity. 
The $c_{\rm AU}$ and $v_{\rm AU}$ values for the two sources are shown in Table \ref{tb:params}, 
where the error ranges for the L1527 case are estimated by eye from the fits with various parameters. 
The dynamical timescales in Table \ref{tb:params} are taken from Y\i ld\i z et al. (2015). 
Since our observations with ALMA are focused on the central region around the protostar, 
the dynamical timescales cannot be estimated adequately from our data. 

Although the inclination angles for the two sources indicate that they both have a nearly edge-on configuration of the outflow/envelope structure, 
the appearance of the outflows shown in their moment 0 maps are quite different. 
The outflow of L1527 shows a butterfly-feature, while IRAS 15398--3359 has a well collimated outflow. 
While we assumed an offset between 
the two lobes in L1527 to explain the outflow structure 
as suggested by Tobin et al. (2008) (Section \ref{ssec:outflow}), 
no such an offset is seen for the two lobes in IRAS 15398--3359. 
It should be noted that the outflow component is contaminated with the envelope component within $1^{\prime \prime}$ from the protostar position in IRAS 15398--3359. 
Apart from the offset, 
the difference of the opening angles of the outflow cavity is reflected in the difference of the values of $C/D$ by a factor of 14 (Table \ref{tb:params}). 
The difference might reflect a different evolutionary stage of the outflow. 

\begin{figure}[h]
	\includegraphics[bb = 0 0 100 300, scale = 1.3]{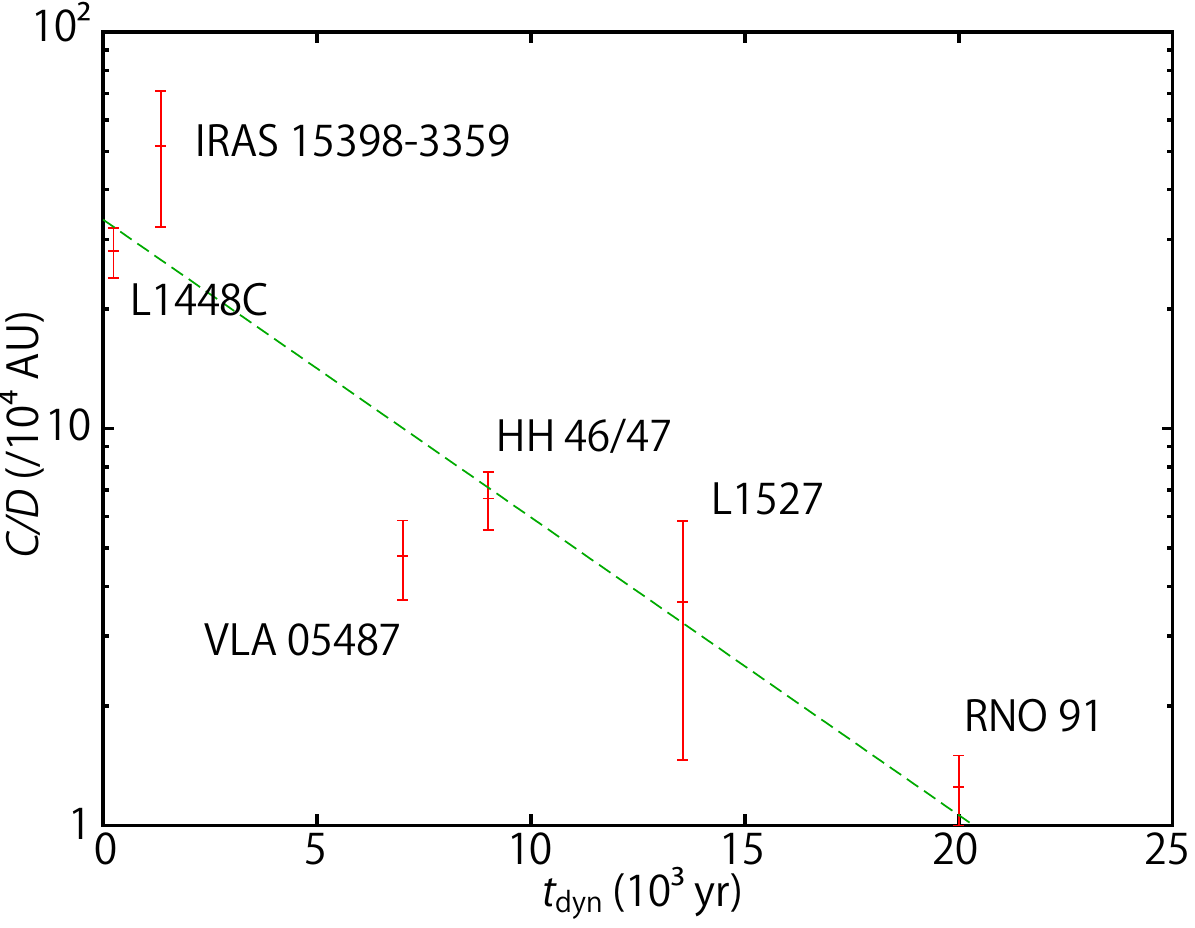}
	\caption{The curvature ($C/D$) of the outflows as a function of the dynamical timescales. 
			The dynamical timescale for L1527 and IRAS 15398--3359 are the averaged value of the two lobes. 
			The green line represents the best-fit function: 
			$\log (C/D) = (-0.17 \pm 0.03) \times (t_{\rm dyn} \times 10^{-3}) + (-5.7 \pm 0.3)$, 
			where the uniform weights are applied to all the sources. 
			The correlation coefficient for this plot is $-0.94$. 
			\label{fig:CDt}}
\end{figure}

The outflow parameters are reported for 
the Class 0/I source VLA 05487 in the Orion dark cloud L1617 
and the Class II/III source RNO 91 in the L43 molecular cloud by Lee et al. (2000). 
Moreover, those for the Class 0 source L1448C (L1448 mm) in Perseus 
are reported by Hirano et al. (2010), 
and those for the HH 46/47 molecular outflow on the outskirts of the Gum Nebula by Arce et al. (2013). 
The outflow parameters for these sources are summarized in Table \ref{tb:params}. 
Here, it should be noted that the coefficient of proportionality of Eq. (\ref{eq:outflow}) (middle), 
$v_0 / R_0$, is given in the above papers. 
Figure \ref{fig:CDt} shows a semi--log plot of $C/D$ versus dynamical timescale $t_{\rm dyn}$ for these six sources. 
The green dotted line in Figure \ref{fig:CDt} represents the best-fit result, 
where an equal weight is assumed for all the sources. 
Although the number of the sources is too small for statistical arguments, 
the parameter $C/D$ seems to decrease exponentially with the dynamical timescale of outflows. 
The correlation coefficient for this plot is $-0.94$. 
This feature is consistent with the previous works reporting a relationship between the opening angles of outflows and the source ages 
(Arce \& Sargent 2006; Seale \& Looney 2008; Velusamy et al. 2014). 
The trend of increasing opening angle as increasing age is also revealed by the theoretical simulations (Offner et al. 2011; Shang et al. 2006). 
However, previous observational studies are based on morphology of outflows. 
Although distribution of outflows is affected by the inclination angle, 
its effect is not considered except for the work by Seale \& Looney (2008). 
In contrast, the outflow parameters reported in the present study are derived from both the geometrical and kinematic structures of the outflow 
by using the parabolic model considering the inclination angle. 
The results quantitatively supports the trend suggested previously (Arce \& Sargent 2006; Seale \& Looney 2008; Velusamy et al. 2014).

\begin{figure}[h]
	\includegraphics[bb = 0 0 100 300, scale = 1.3]{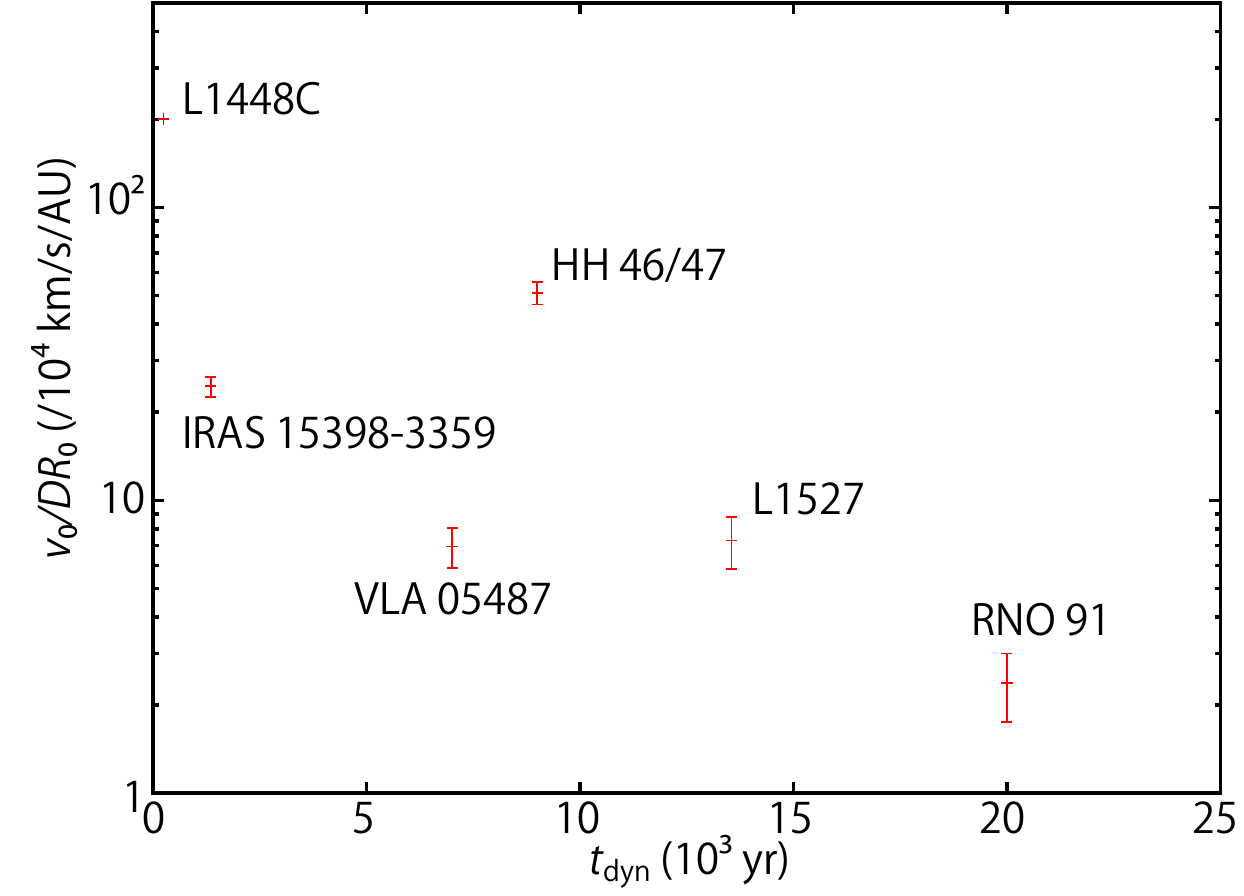}
	\caption{The velocity parameter ($v_0/DR_0$) of the outflows as a function of the dynamical timescale. 
			The dynamical timescales for L1527 and IRAS 15398--3359 are the averaged value of the two lobes. 
			The correlation coefficient for this plot is $-0.80$, 
			where the uniform weights are applied to all the sources. 
			\label{fig:v0Dt}}
\end{figure}

Figure \ref{fig:v0Dt} shows a semi--log plot of $v_0/DR_0$ versus dynamical timescale $t_{\rm dyn}$. 
Although the correlation coefficient of $-0.80$ is lower than that for the plot in Figure \ref{fig:CDt}, 
the velocity of the gas also seems to decrease exponentially with the dynamical timescale of outflows. 
There seems to be a trend that 
outflows with shorter dynamical timescales have higher 
$v_0 / D R_0$. 
This trend is derived from the velocity structure of the outflow, 
which was not considered in the morphological studies (Arce \& Sargent 2006; Seale \& Looney 2008; Velusamy et al. 2014). 
Although the parameters for L1527 are obtained with an observation focused on a narrow region without the whole structure of the outflow, 
they are consistent with the values reported previously by the analysis of the large-scale outflow structures. 
The outflow can be characterized by using a parabolic outflow model focused on a narrow region around the protostar, 
even if the whole structure is not observed. 
This strengthens the idea that the acceleration and geometry of the outflow are mostly defined 
in a very small region near to the source. 

In this paper, we investigated the kinematic structure of the outflow in L1527 with the aid of the kinematic structure of the envelope. 
Recently, spatial resolution in millimeter/submillimeter-wave observations has become much higher with ALMA than before, and hence, 
it will be possible to unveil the small-scale structure of the envelope, 
especially in the vicinity of the centrifugal barrier. 
In addition to the envelope, 
we here stress that the outflow can be investigated from observations focused on a narrow region around the protostar by use of the outflow model. 
Outflow morphology in a large scale is often affected by interactions with the ambient gas, 
and elimination of this effect is indispensable for characterization of outflows, 
as discussed by Velusamy et al. (2014). 
Hence, observations of outflows in the vicinity of the protostar 
will be an important technique in the further outflow studies, 
as demonstrated in this study. 
Moreover, such high-resolution observations will delineate both the inner structure of the envelope and the launching point of the outflow, 
which would be related to each other in forming the rotationally-supported disks. 

\acknowledgments
The authors are grateful to John Tobin and Yasushi Suto for their invaluable discussions. 
This paper makes use of the following ALMA data set ADS/JAO.ALMA$\#$2011.0.00604.S. 
ALMA is a partnership of the ESO (representing its member states), 
the NSF (USA) and NINS (Japan), together with the NRC (Canada) and the NSC and ASIAA (Taiwan), 
in cooperation with the Republic of Chile. 
The Joint ALMA Observatory is operated by the ESO, the AUI/NRAO and the NAOJ. 
The authors are grateful to the ALMA staff for their excellent support. 
Y.O. acknowledges financial support from the Advanced Leading Graduate Course for Photon Science (ALPS). 
Y.O. also acknowledges the JSPS fellowship. 
This study is supported by Grant-in-Aid from the Ministry of 
Education, Culture, Sports, Science, and Technologies of Japan (21224002, 25400223, 25108005, and 15J01610). 
N.S. and S.Y. acknowledge financial support by JSPS and MAEE under the Japan-France integrated action program (SAKURA: 25765VC).

\appendix
\section*{Appendix}
\section{Envelope model with various inclination angles}
Figure \ref{fig:PV_Idiff} shows the PV diagrams of CS ($J$=5--4; color) along the two lines 
passing through the protostar position shown in Figure \ref{fig:PV_direction}: 
one is perpendicular to the outflow axis (``$180\degr$") and the other is along it (``$270\degr$"). 
The blue contours represent the infalling rotating envelope models with various inclination angles. 
In these models, the protostellar mass and the radius of the centrifugal barrier are assumed to be 0.18 $M_\odot$ and 100 AU, respectively, 
which are derived by Sakai et al. (2014b). 
The outer radius of the envelope is fixed to be 1000 AU, which well reproduces the PV diagram of CCH (Sakai et al. 2014a). 
The density profile is assumed to be proportional to $r^{-1.5}$. 
The spectral line is assumed to be optically thin, 
and to have a Gaussian profile with the line width of 0.5 km s$^{-1}$. 
The emission is convolved with a Gaussian beam with FWHM of $0\farcs5 \times 0\farcs5$. 

The optically thin condition for the velocity components outside the systemic velocity (5.9 km s$^{-1}$; Sakai et al. 2010) can be justified for the CS line, 
because the optical depth toward the centrifugal barrier estimated by the LVG analysis is 0.1 
for the H$_2$ density range form $3 \times 10^6$ to $3 \times 10^7$ cm$^{-3}$ 
and the temperature range from 30 to 60 K. 
Although the CS line is heavily self-absorbed at the systemic velocity, 
we do not consider this component in comparison with the models. 

\begin{figure}[h]
	\includegraphics[bb = 0 0 100 1000, scale = 0.5]{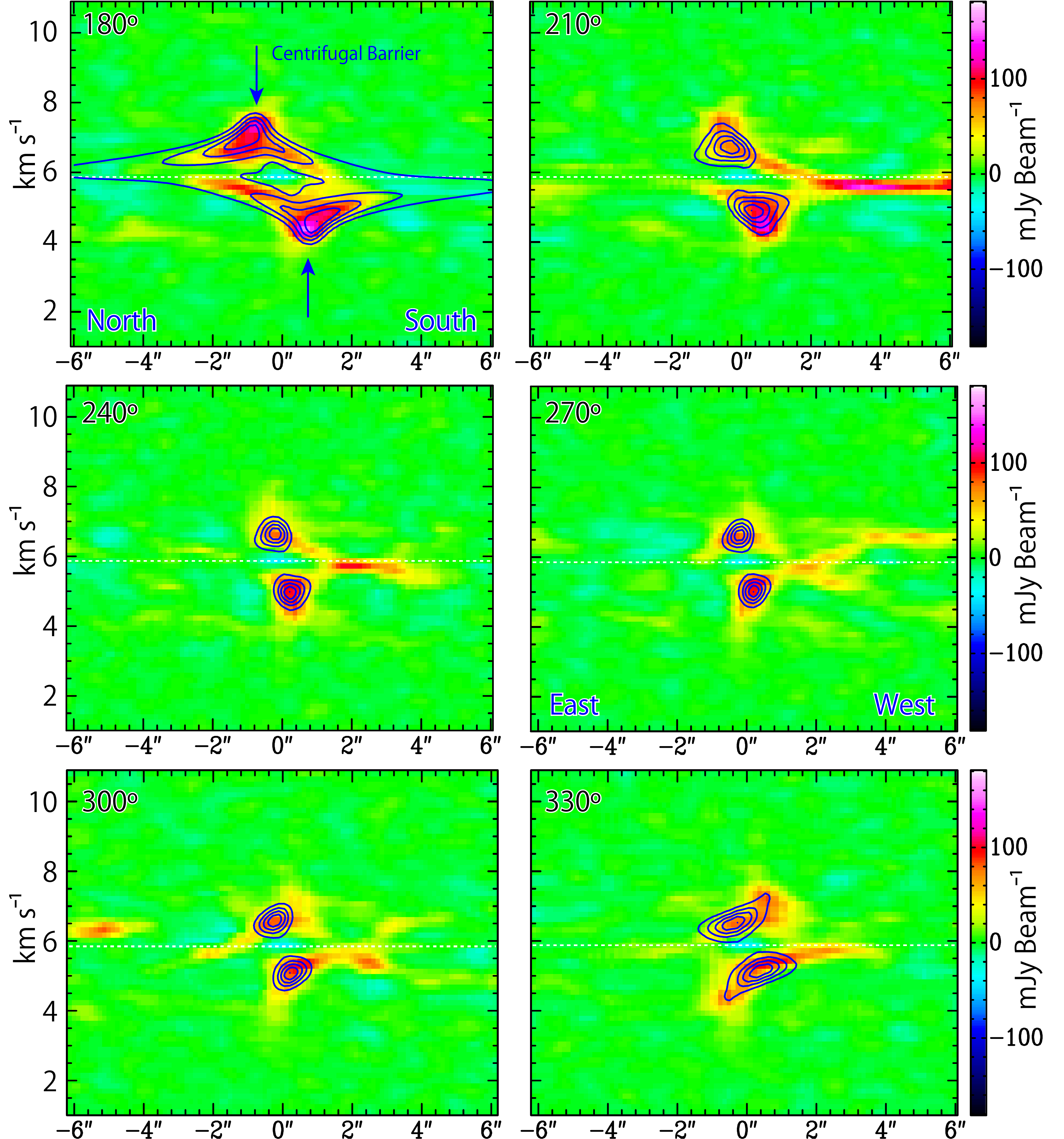}
	\caption{PV diagrams of CS ($J$=5--4; color) along the lines shown in Figure \ref{fig:PV_direction}(a). 
			The blue contours represent the infalling rotating envelope model with the inclination angle of $+5\degr$ (Figure \ref{fig:PV_direction}(b)). 
			Contours are every 20 \% of the peak intensity. 
			\label{fig:PV_I5}}
\end{figure}

\begin{figure}[h]
	\includegraphics[bb = 0 0 100 1000, scale = 0.5]{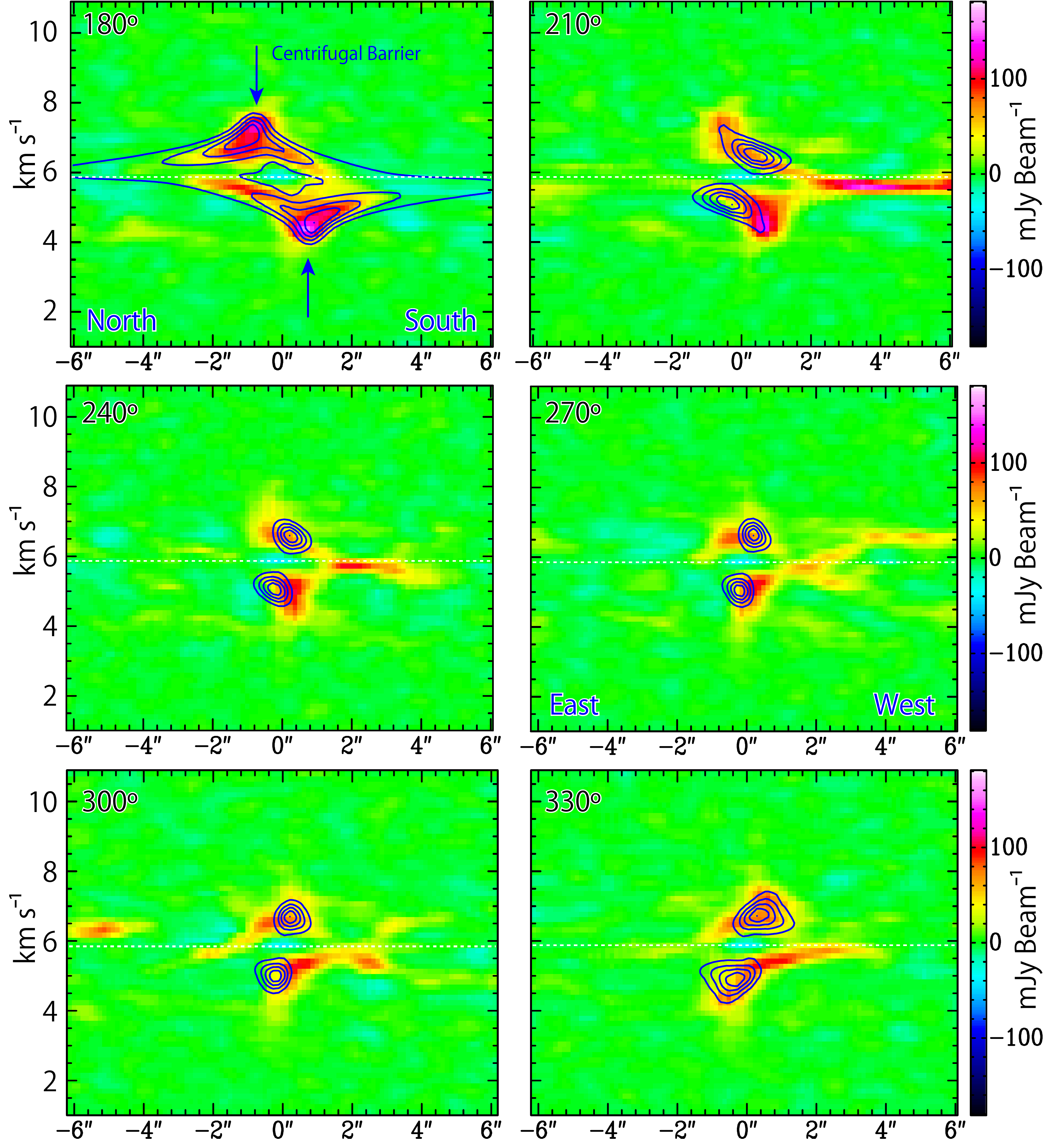}
	\caption{PV diagrams of CS ($J$=5--4; color) along the lines shown in Figure \ref{fig:PV_direction}(a). 
			The blue contours represent the infalling rotating envelope model with the inclination angle of $-5\degr$ (Figure \ref{fig:PV_direction}(c)). 
			Contours are every 20 \% of the peak intensity. 
			\label{fig:PV_I-5}}
\end{figure}

\begin{figure}[h]
	\includegraphics[bb = 0 0 100 500, scale = 0.45]{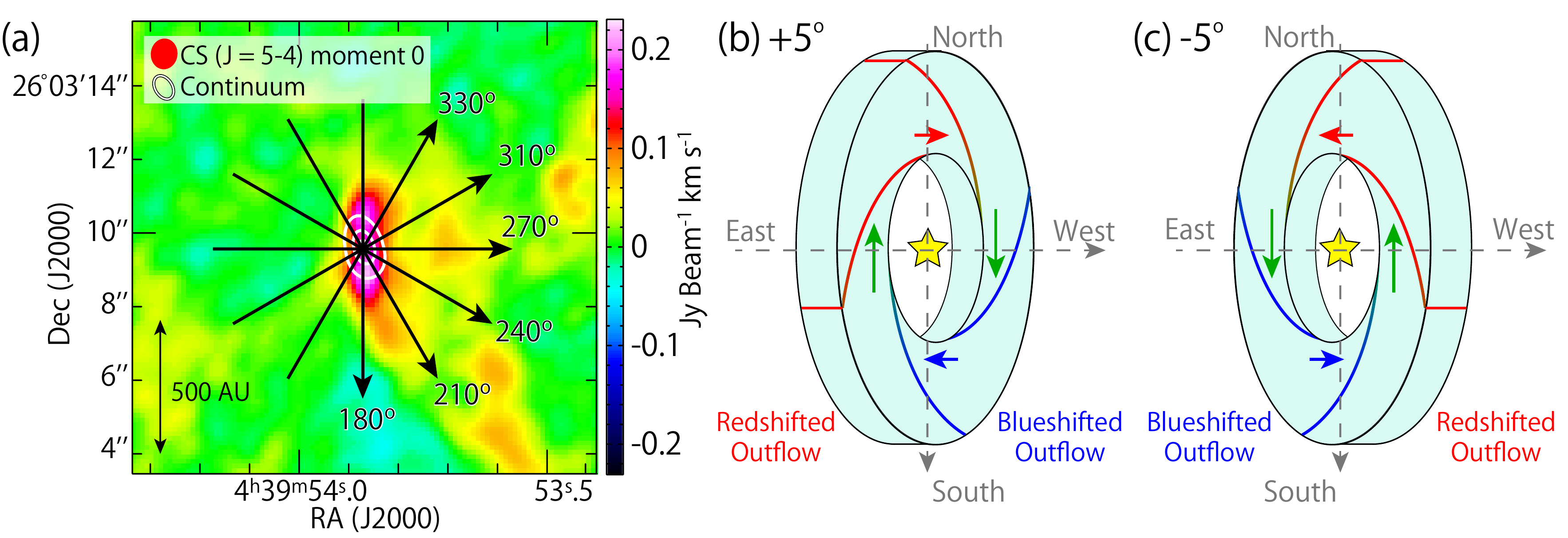}
	\caption{(a) A blow-up of the central part of the moment 0 map of CS ($J$=5--4; color) and the continuum map (white contours) shown in Figure \ref{fig:moment0}. 
			Contours for continuum are as the same as in Figure \ref{fig:moment0}. 
			The black arrows represent the lines along which the PV diagrams 
			in Figures \ref{fig:PV_I5}, \ref{fig:PV_I-5} and \ref{fig:PV_c-C3H2} are prepared. 
			(b) Schematic illustration of the envelope model with the inclination angle of $+5\degr$. 
			(c) Schematic illustration of the envelope model with the inclination angle of $-5\degr$. 
			\label{fig:PV_direction}}
\end{figure}

\begin{figure}[h]
	\includegraphics[bb = 0 0 100 1000, scale = 0.5]{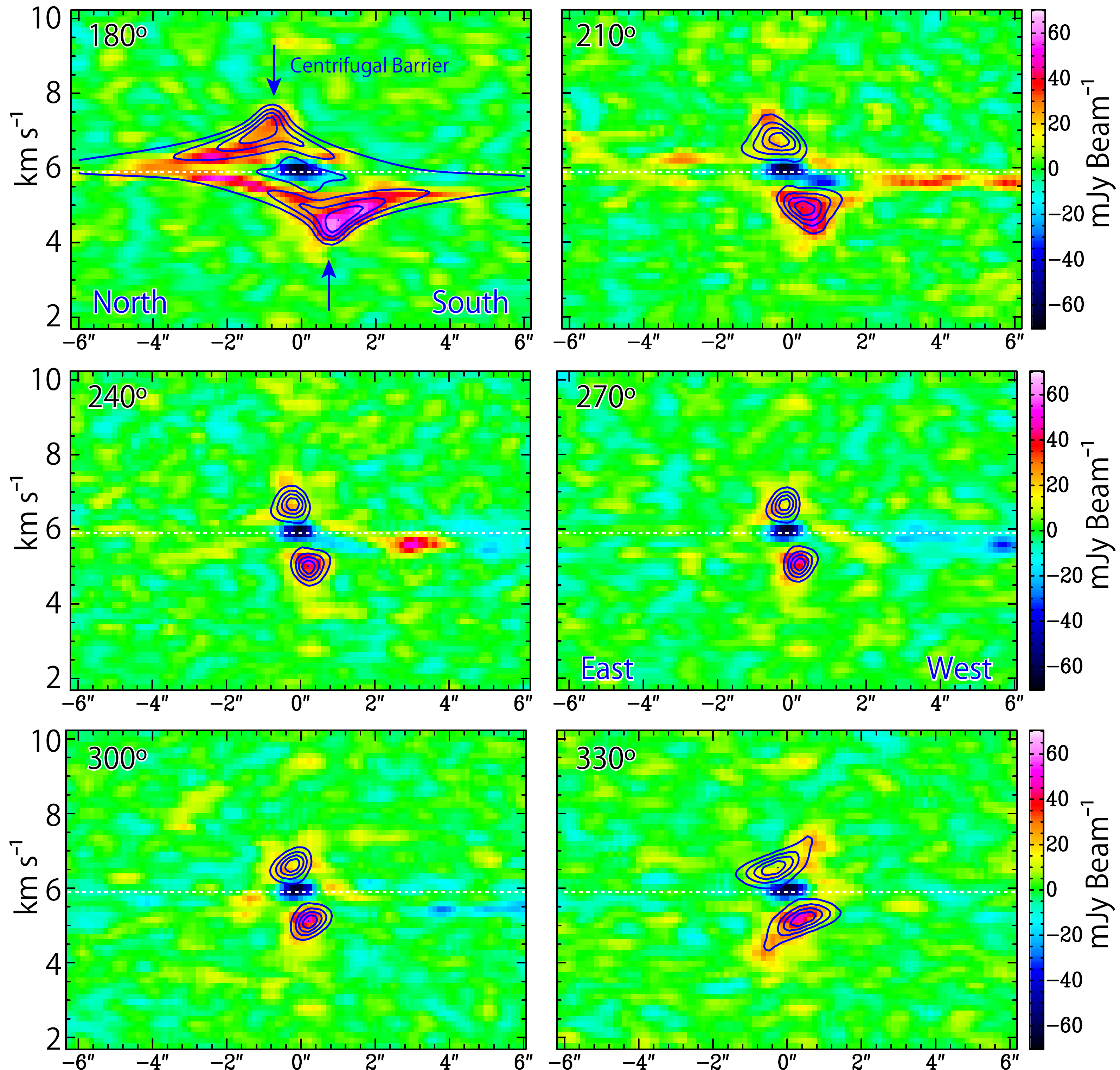}
	\caption{PV diagrams of c-C$_3$H$_2$ ($5_{2, 3}$--$4_{3, 2}$; color) along the lines shown in Figure \ref{fig:PV_direction}(a). 
			The blue contours represent the infalling rotating envelope model with the inclination angle of $+5\degr$. 
			Contours are every 20 \% of the peak intensity. 
			\label{fig:PV_c-C3H2}}
\end{figure}

\begin{figure}[h]
	\includegraphics[bb = 0 0 100 1000, scale = 0.52]{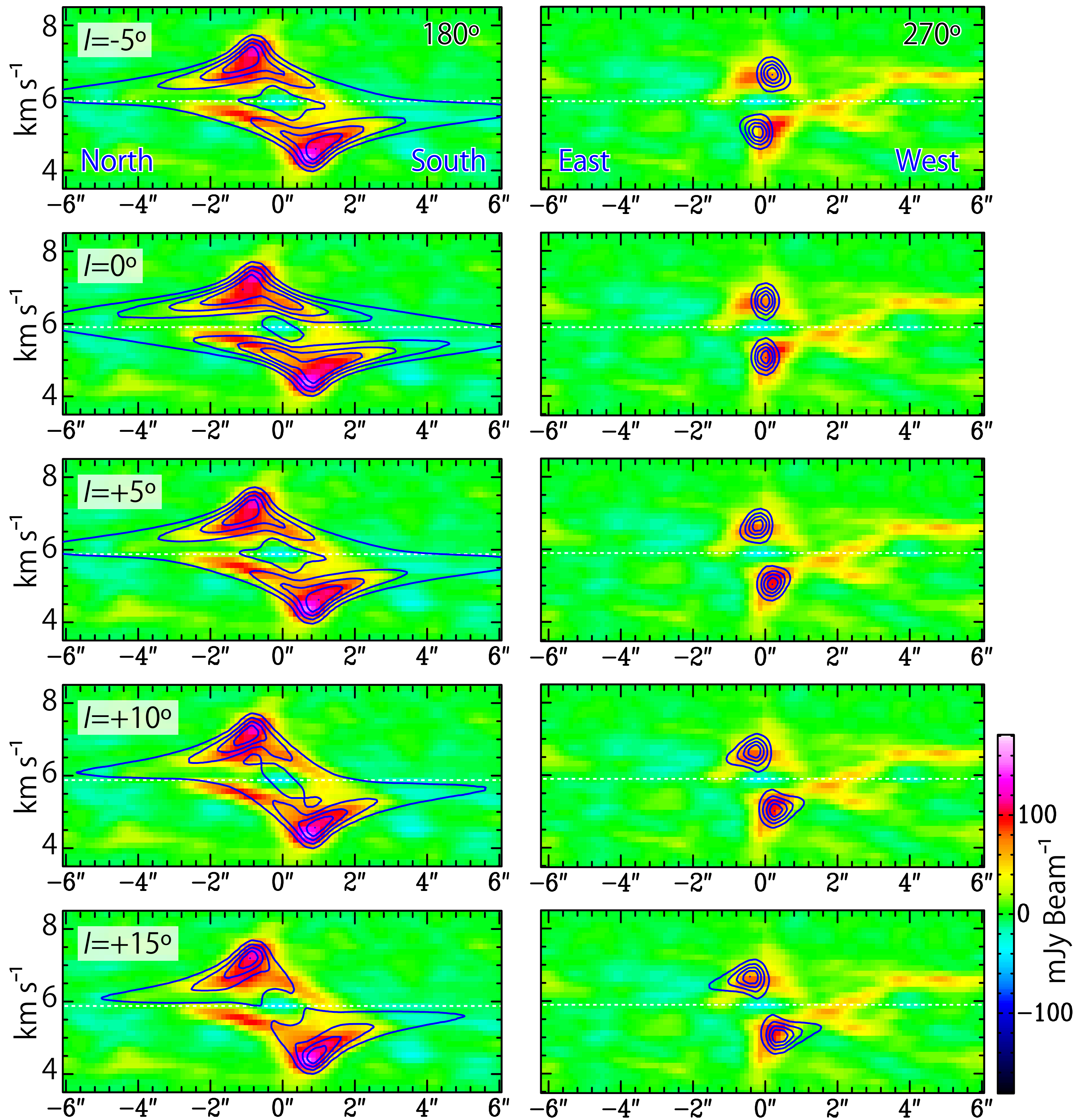}
	\caption{PV diagrams of CS ($J$=5--4; color) along the lines shown in Figure \ref{fig:PV_direction}(a). 
			The blue contours in the two diagrams in each row represent the infalling rotating envelope model with the inclination angle of $-5\degr$ (Figure \ref{fig:PV_direction}(c)), $0\degr$ (edge-on), $+5\degr$ (Figure \ref{fig:PV_direction}(b)), $+10\degr$, or $+15\degr$. 
			Contours are every 20 \% of the peak intensity. 
			The panels for the inclination angles of $+5\degr$ and $-5\degr$ are 
			the same as those presented in Figures \ref{fig:PV_I5} and \ref{fig:PV_I-5}, respectively. 
			\label{fig:PV_Idiff}}
\end{figure}

Panels for the inclination angles of $+5\degr$ and $-5\degr$ in Figure \ref{fig:PV_Idiff} are the same as 
the corresponding panels in Figures \ref{fig:PV_I5} and \ref{fig:PV_I-5}, respectively. 
As discussed in Section \ref{ssec:env}, the PV diagram along the line perpendicular to the outflow axis (i.e. along the envelope; ``$180\degr$") reflects 
the absolute value of the inclination angle regardless of the directions of the inclination. 
On the other hand, the PV diagram along the outflow axis (``$270\degr$") reflects the infalling motion of the envelope more significantly, 
and is sensitive to the direction of the inclination. 


\clearpage
\clearpage
\clearpage
\clearpage
\clearpage
\clearpage
\clearpage
\clearpage
\clearpage
\clearpage
\clearpage
\clearpage

\clearpage

\clearpage
\begin{table}
	\caption{Parameters of the Observed Line\tablenotemark{a} \label{tb:line}}
	\begin{center}
		\begin{tabular}{llccc}
			\hline \hline 
			Molecule & Transition & Frequency & $E_{\rm u}$ & S$\mu^2\ \tablenotemark{b}$ \\ 
			 & & (GHz) & (K) & (D$^2$) \\ 
			 \hline
			 CS & $J$=5--4 & 244.9355565 & 35 & 19.17 \\ 
			 c-C$_3$H$_2$ & $5_{2, 3}$--$4_{3, 2}$ & 249.0543680 & 41 & 76.32 \\ 
			 \hline 
		\end{tabular}
	\end{center}
	\tablenotetext{a}{Taken from CDMS (M\"{u}ller et al. 2005).}
	\tablenotetext{b}{Nuclear spin degeneracy is not included.}
\end{table}

\clearpage
\begin{landscape}
\begin{table}
	\caption{Best fit parameters for the models of the outflow cavity walls and the dynamical timescales \label{tb:params}}
	\begin{center}
		\begin{tabular}{lccccc}
			\hline \hline
			Source & Distance & $t_{\rm dyn}$ & Inclination Angle & $c_{\rm AU} = C/D$ & $v_{\rm AU} / r_0 = v_0 / D R_0$ \\ 
			& (pc) & ($10^3$ yr) & ($\degr$) & ($10^{-4}$ AU$^{-1}$) & ($10^{-4}$ km s$^{-1}$ AU$^{-1}$) \\ 
			\hline 
			L1527 & 137 & 20.6 (East), 6.5 (West) \tablenotemark{a} & 5 $\pm$ 10 & 3.6 $\pm$ 2.2 & 7.3 $\pm$ 1.5 \\ 
			IRAS 15398--3359\tablenotemark{b} & 155 & 0.9 (Red), 1.8 (Blue)\tablenotemark{a} & 20 $\pm$ 10 & 52 $\pm$ 19 & 25 $\pm$ 2 \\ 
			VLA 05487\tablenotemark{c} & 460 & $\sim$7& 19 $\pm$ 3 & 4.8 $\pm$ 1.1 & 7.0 $\pm$ 1.1 \\ 
			RNO 91\tablenotemark{c} & 160 & $\sim$20 & 20 $\pm$ 4 & 1.3 $\pm$ 0.25 & 2.4 $\pm$ 0.6 \\ 
			L1448C\tablenotemark{d} & 250 & $\sim 0.24$& 21 & 24, 32 & 200 \\ 
			HH 46/47\tablenotemark{e} & 450 & 9 & 29 $\pm$ 1 & 6.7 $\pm$ 1.1 & 51 $\pm$ 4 \\ 
			\hline 
		\end{tabular}
	\end{center}
	\tablenotetext{a}{Determined from the CO ($J$=3--2) emission (Y\i ld\i z et al. 2015). }
	\tablenotetext{b}{Determined from the H$_2$CO ($5_{1, 5}$--$4_{1, 4}$) emission (Oya et al. 2014). }
	\tablenotetext{c}{Determined from the CO ($J$=1--0) emission (Lee et al. 2000). }
	\tablenotetext{d}{Determined from the CO ($J$=3--2) emission (Hirano et al. 2010). }
	\tablenotetext{e}{Determined from the CO ($J$=1--0) emission (Arce et al. 2013). }
\end{table}
\end{landscape}

\end{document}